\documentclass[onecolumn,12pt]{IEEEtran}
\usepackage[square,sort&compress,comma,numbers]{natbib}
\usepackage{color}
\usepackage{graphicx}
\usepackage{amsfonts}
\usepackage{epstopdf}
\usepackage{latexsym}
\usepackage{amssymb}
\usepackage{amsmath}
\usepackage{fixltx2e}

\ifCLASSOPTIONcompsoc

	\usepackage[caption=false, font=normalsize, labelfont=sf, textfont=sf]{subfig}
	\else
	\usepackage[caption=false, font=footnotesize]{subfig}
\fi

\begin{document}

\title{Study on Compressed Sensing of Action Potential}

\author{\IEEEauthorblockN
    {Hyunseok~Park, Xilin~Liu
}
\IEEEauthorblockA{\\Department of Electrical and Systems Engineering (ESE)\\
University of Pennsylvania, Philadelphia, PA 19104} }
\maketitle

\begin{abstract}
Compressive sensing (CS) is a signal processing
technique that enables sub-Nyquist sampling and near lossless reconstruction
of a sparse signal. The technique is particularly
appealing for neural signal processing since it avoids the
issues relevant to high sampling rate and large data storage.
In this project, different CS reconstruction algorithms were
tested on raw action potential signals recorded in our lab.
Two numerical criteria were
set to evaluate the performance of different CS algorithms:
Compression Ratio (CR) and Signal-to-Noise Ratio (SNR). In
order to do this, individual CS algorithm testing platforms
for the EEG data were constructed within MATLAB scheme.
The main considerations for the project were the following.
1) Feasibility of the dictionary 2) Tolerance to non-sparsity
3) Applicability of thresholding or interpolation.
\end{abstract}

\begin{IEEEkeywords}
Compressive Sesning (CS), Nyquist-Shannon
Sampling, Electrocephalography (EEG), Sparsity
\end{IEEEkeywords}

\IEEEpeerreviewmaketitle

\section{Introduction}

Electrophysiological signals present brain activities in the form of electrical signals. Action potential is the activity of single neuron, which consists of rapid polarization and depolarization process.
Action potential is key in understanding neuron activities and brain-machine interface applications. Action potential is typically recorded at ten of kilohertz [1].
Unfortunately, recording action potentials from multiple neurons pose significant power penalty [2].
In order to conquer the power limitation, we study the use of compressive sensing (CS) for acquiring action potential signals.

Unlike traditional sampling-to-compression method, CS takes advantage of the sparsity of the signal [3]. Ideally, the sampled data can be near losslessly reconstructed [4]. Yet, reconstruction results
critically depends on the sparsity of the signal of interest [5].
If a signal fails to be sparse in any representable domain, a
promising reconstruction cannot be obtained.

Using the CVX tool and the BSBL-BO CS algorithm on MATLAB, seven distinct domains were tested for CS reconstruction results. The implementation was successful, however, the achievalb compression ratio is limited.
Although the reconstruction result may be further improved with better sprase dictionary, we conclude CS is not ideal for compressing action potential signals. Hardware sensing implementation is left for the future work with the random matrix prepared in this experiment.

\section{Action Potential Data}
The dataset used for this work was collected in our lab [6-7]. The original sampling rate is 15kSps. An analog gain of 60dB is applied when acquiring the data. A sample length of N = 256 was used, as shown in Fig. 1.
\begin{figure}
  \centering
  \includegraphics[]{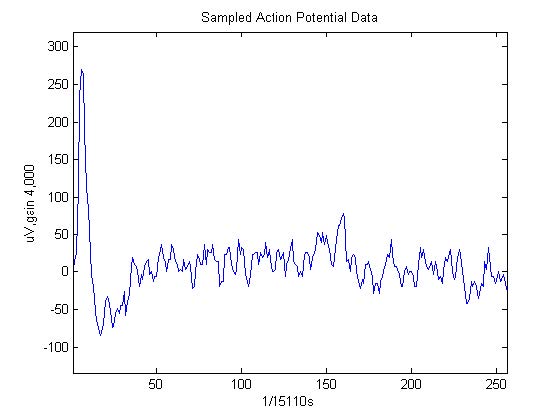}
  \caption{An example action potential data vector with a length of 256 samples. }\label{fig1}
\end{figure}
The sampled signal vector is composed of single spiking greater than 250$\mu$V and noise limited by thermal and instrumentation used.
Without loss of generality, N, M, and different segment of sample can be selected for the experiment with desired compression ration (CR) [8].

\section{EXPERIMENTAL RESULTS}

\subsection{CVX Sparse-Dictionary CS Result}

On MATLAB, CVX convex optimization solver tool was
installed to solve the reconstruction.
The sampling matrix was constructed with normally distributed
pseudorandom numbers. To retain consistency of the testing,
The sampling matrix was fixed.
Time domain reconstruction result is shown in Figure 2-(a). The result shows weak reconstruction due to nonsparsity
of the signal in time domain, as expected. Figure 2-(b)
illustrates reconstruction result with discrete cosine transform (DCT)
matrix. The matrix was constructed by transforming an identity
matrix N $\times$ N.

\begin{figure*}[!ht]
  \centering
  \includegraphics[width = 1\textwidth]{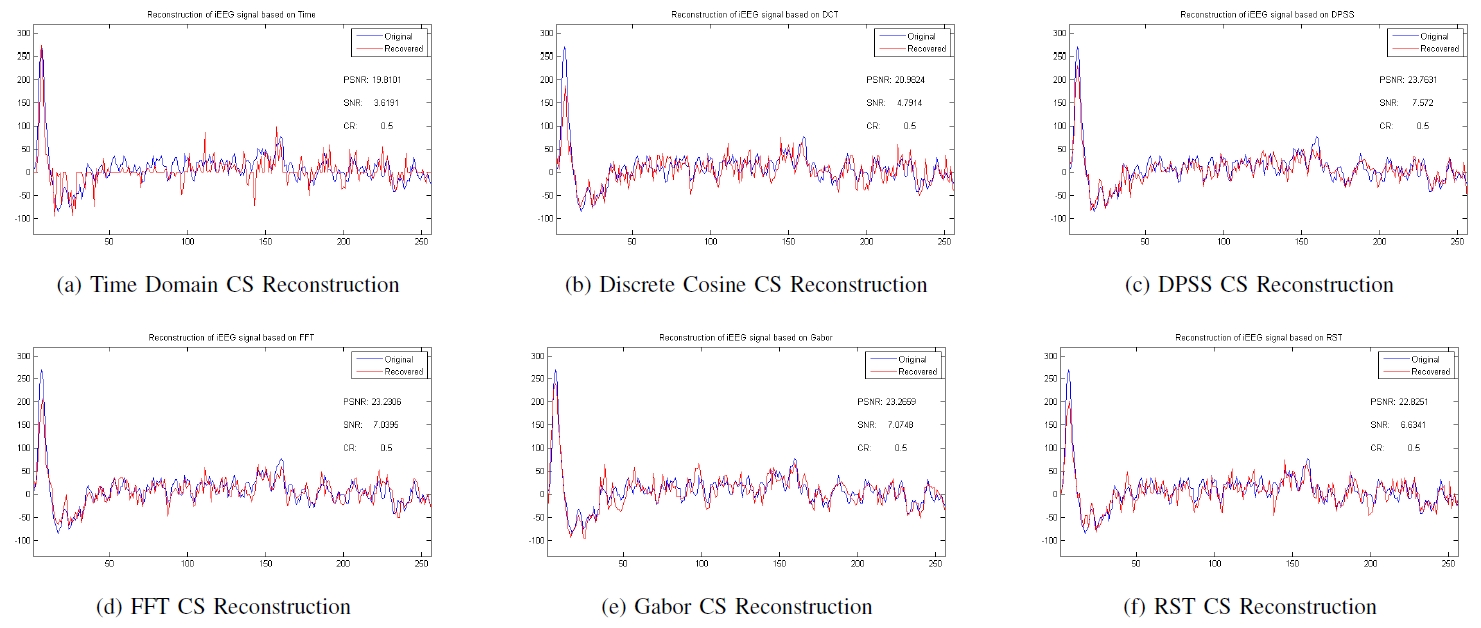}
  \caption{The results from compressive sensing reconstruction on 6 different sparse-basis. }\label{fig1}
\end{figure*}

The sparse-matrix used in the reconstruction in Figure 2-(c)
was composed of discrete prolate spheroidal sequence (DPSS)
with time-half bandwidth set to 8.5. DPSS forms highly
efficient basis for sampled bandlimited functions, compared
to conventional fourier series [9]. Figure 2-(d) illustrates the
reconstruction result with sparse-matrix formed by fast fourier
transform (FFT).

Only real parts of the result was extracted for the reconstruction
vector xrecons with magnitude of the imaginary
components. Gabor basis for 2-(e) was formed with
sine and cosine basis with Gaussian kernels. The dictionary
functions suggested by Abdulghani et al., where the parameters n is the sample size, n0 is the sample number of the centre of the envelop, $\omega >$ 0 is the sinusoid’s
frequency, $\sigma$ > 0 is the spread of the envelope, and $\theta$
is the phase angle [10]. Noiselet dictionary, composed of
functions complementary to wavelet, was constructed with
the generation code available, constructed by Laurent Duval.
And lastly, real sinusoid transform (RST) was performed on
identity matrix using SparseLab2.1-Core tool.

Operation time wise, RST showed the best performance
with 2.48s followed by DPSS with 2.73s. Both SNR and
time wise, DPSS illustrated the overall superior performance.

\subsection{BSBL-BO Sparse Dictionary CS Result}

It has been verified in the pervious section that basis pursuit
with CVX tool failed to give aimed result, SNR $>$ 10dB
at CR = 0:5. This is due to inherent non-sparsity of the
tested EEG signal. A CS reconstruction algorithm that is less
sensitive to the non-sparsity was necessary. The algorithm
Block Sparsity Bayesian Learning-Bound Optimization, or
BSBL-BO by Zhilin Zhang is tolerant to the non-sparsity of
the signal while successfully reconstructing the signal using
block sparsity.

With block-partition
sparsity and intra-block correlation, the algorithm itself is less
sensitive to the choice of sparse-dictionary. That is, although
the performance still depends on the sparse-dictionary, the
reconstruction result is overall better than CVX results. The
same procedure was followed for testing BSBL-BO with
time domain replaced by Wavelet of Daubechies-20, using
WaveLab850.

\begin{figure*}[!ht]
  \centering
  \includegraphics[width = 1\textwidth]{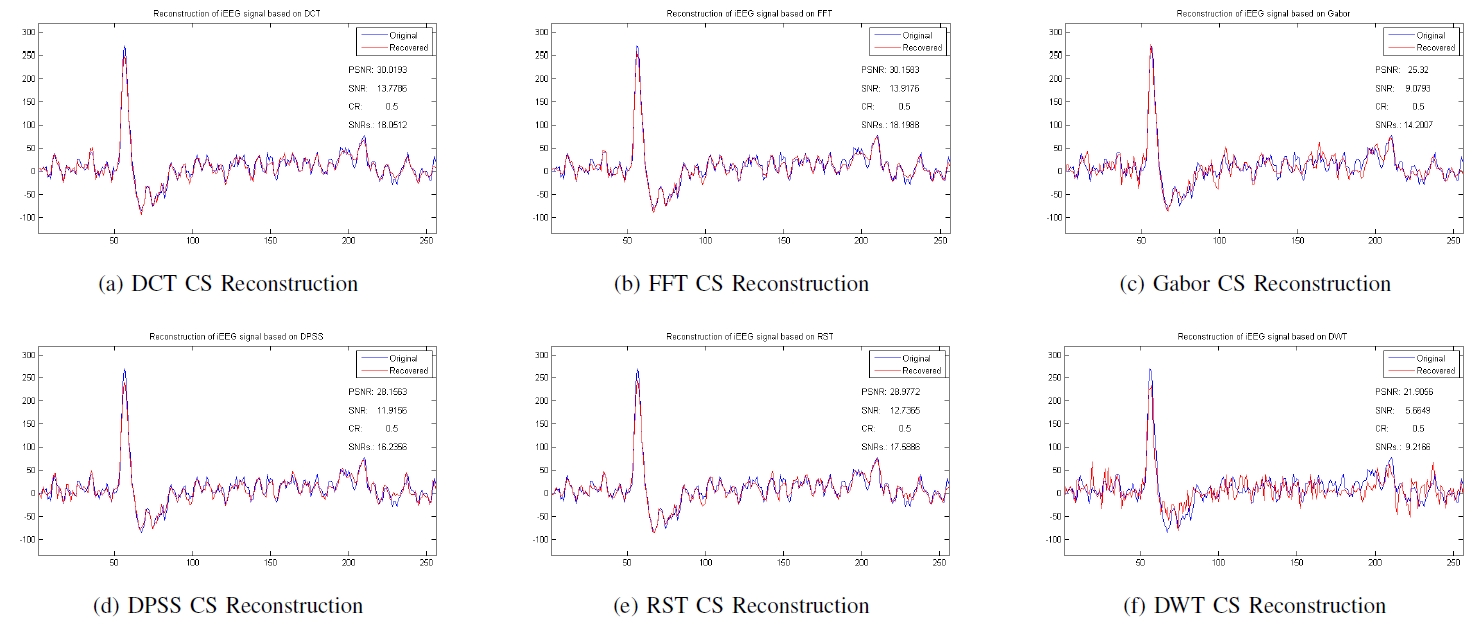}
  \caption{CS reconstruction was tested using BSBL-BO algorithm with variation to sparse-basis. }\label{fig1}
\end{figure*}

Refer to figure 3, All the reconstruction results except for
noislet, gabor, and wavelet basis show SNR level$>$10dB
for the aimed reconstruction result. The algorithm execution
time is optimized and overall faster than CVX solver.
Fourier basis illustrated the best performance in PSNR with
30.15dB and SNR with 13.91dB. DCT demonstrated
the shortest operation time with 0.075s, which shows marginal
difference of 0.003s compared to FFT. In
short, sparse-dictionary composed of fourier series showed
superior performance for BSBL-BO algorithm.

The window size of 30 or the total size of 60 was set with respect to the spike on the center.
The SNR was derived from that window only, that is, the spike
SNR was calculated. The purpose was to verify performance
of the reconstruction of spikes that likely contain information
related to neuron firing activity. The basis comparison results
are presented in Table 1.

Further testing was performed by varying CR with
Fourier sparse matrix and fixed sensing matrix A. The code
constructed by Igor Carronw as used to construct sparse
sampling matrix that composed of only binary number entries. Graphical
results are presented in Figure 4.

\begin{table}[!ht]
\centering
\caption{BSBL-BO DICTIONARY PERFORMANCE TABLE}
\begin{tabular}{|c|c|c|c|c|}
\hline
Basis           & PSNR(dB) & SNR(dB) & CR  & Time (s) \\ \hline
Discrete Cosine & 30.0193  & 13.7786 & 0.5 & 0.075    \\ \hline
Fourier         & 30.1583  & 13.9176 & 0.5 & 0.182    \\ \hline
Gabor           & 25.32    & 9.0793  & 0.5 & 0.078    \\ \hline
DPSS            & 28.1563  & 11.9156 & 0.5 & 0.078    \\ \hline
Noiselet        & 21.0742  & 4.8335  & 0.5 & 0.201    \\ \hline
Real Sinusoidal & 28.9772  & 12.7365 & 0.5 & 0.086    \\ \hline
Wavelet         & 21.9056  & 5.6649  & 0.5 & 0.303    \\ \hline
\end{tabular}
\end{table}

Interpolation was tested with a window size of 30 with respect to the
spike centre, the reconstruction was tested. The sampling matrix was fixed with
random binary number matrix and Fourier basis was used for
sparse-matrix on BSBL-BO. The data length of N = 60 was
extracted. However, the lack of data points illustrated rugged
signal representation, so the data was interpolated to generate
values inbetween each integer data points. The results were
PSNR of 14.64 and SNR of 3.53 with a CR of 0.5.

However, the result was worse than the N $>$ window:size
reconstruction in the previous sections. Thus, interpolation
method failed to give aimed result.

\begin{figure}[!ht]
  \centering
  \includegraphics[width = 0.7\textwidth]{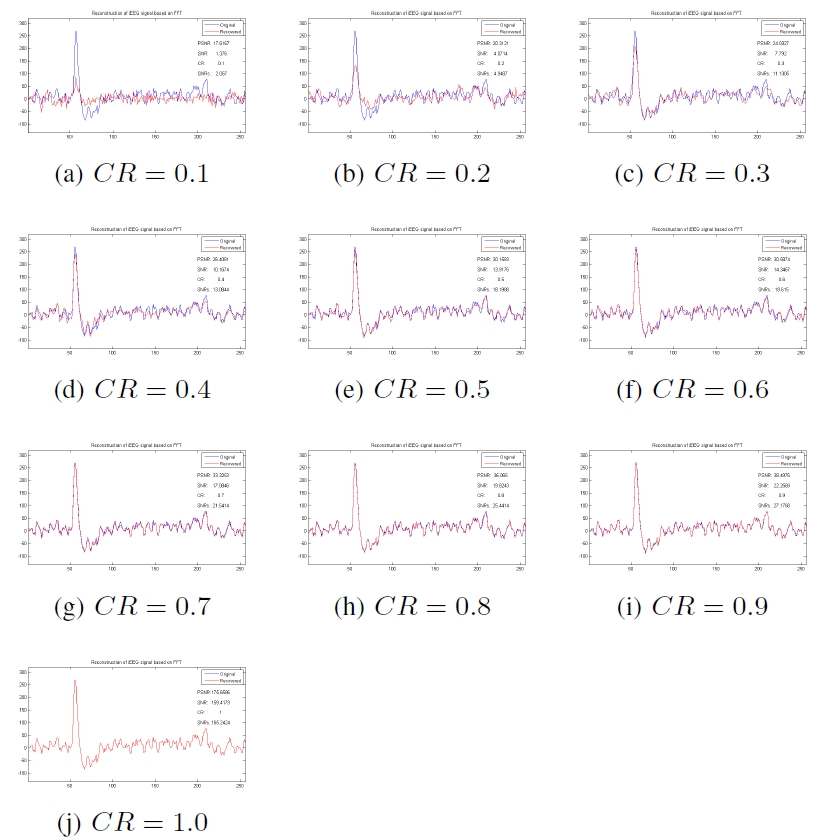}
  \caption{BSBL-BO CS reconstruction results using FFT dictionary were
recorded. Variations to the number of observations M and accordingly CR
were made.}\label{fig1}
\end{figure}

\begin{figure}[!ht]
  \centering
  \includegraphics[width = 0.6\textwidth]{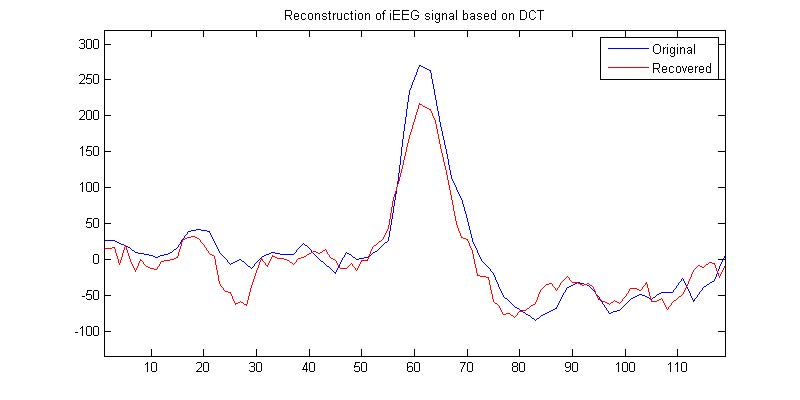}
  \caption{BSBL-BO CS reconstruction results using FFT dictionary were
recorded. Variations to the number of observations M and accordingly CR
were made.}\label{fig1}
\end{figure}

\section{DISCUSSION}

In this project, two different algorithms were used to test
CS reconstruction performance. For CVX
tool based CS reconstruction, discrete prolate spheroidal sequence
basis showed a satisfactory result. For Block
Sparsity Bayesian Learning-Bound Optimization (BSBL-BO)
CS reconstruction, Fourier basis achieved a satisfactory result.
Algorithm-wise, BSBL-BO demonstrated a superior overall performance,
considering SNR-to-CR performances and
algorithm execution time.

Figure 6 shows the SNR-to-CR performance
comparison result. In this project, BSBL-BO was the strongest algorithm candidate for our future
work, considering the inherent
non-sparsity of the signal of interest. However, a higher compression ratio was not achievable in this work. CS may or may not be suitable for compressing action potential signals, depending on the target applications and the hardware resources.

\begin{figure}[!ht]
  \centering
  \includegraphics[width = 0.6\textwidth]{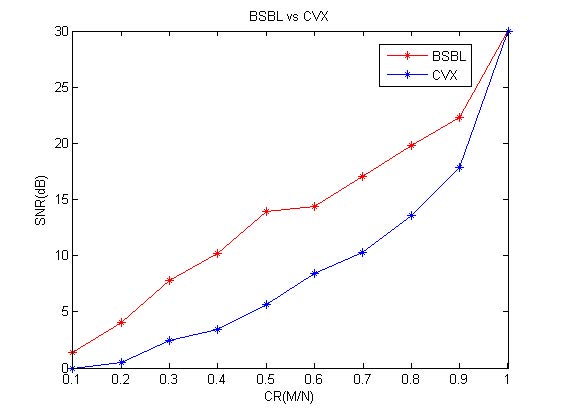}
  \caption{The CS reconstruction results from BSBL-BO and CVX were
compared throughout different CR. For CR = 1, SNR was fixed at 30dB
to facilitate comparison.}\label{fig1}
\end{figure}

\ifCLASSOPTIONcaptionsoff
  \newpage
\fi


\normalsize
\begin{thebibliography}{99}
\footnotesize

\bibitem{3}
M. Zhang, Z. Tang, X. Liu, and J. Van Der Spiegel, “Electronic Neural Interfaces,” Nat. Electron., pp. 1–13, 2020.

\bibitem{PennBMBI2014}
X. Liu, M. Zhang, A. G. Richardson, T. H. Lucas and J. Van der Spiegel, ··A 12-channel bidirectional neural interface chip with integrated channel-level feature extraction and PID controller for closed-loop operation," IEEE Biomedical Circuits and Systems Conference (BioCAS), 2015.

\bibitem{a1}
R. Baraniuk, “Compressive sensing,” IEEE Signal Process. Mag., no. July, pp. 118–121, 2007.

\bibitem{2}
M. B. Wakin, “An Introduction To Compressive Sampling,” IEEE Signal Process. Mag., no. March 2008, pp. 21–30, 2008.


\bibitem{4}
K.-L. Du, M. N. S. Swamy, K.-L. Du, and M. N. S. Swamy, “Compressed Sensing and Dictionary Learning,” Neural Networks Stat. Learn., vol. 73, pp. 525–547, 2019.

\bibitem{5}
X. Liu, M. Zhang, T. Xiong, et al., “A Fully Integrated Wireless Compressed Sensing Neural Signal Acquisition System for Chronic Recording and Brain Machine Interface,” TBioCAS, vol. 10, no. 4, pp. 1–10, 2016.


\bibitem{Spinger2017}
X. Liu, et al., ``Brain-Machine Interface - Closed-loop Bidirectional System Design," \emph{Springer}, Oct. 2017.


\bibitem{7}
A. Dixon, “Compressed sensing system considerations for ECG and EMG wireless biosensors,” TBioCAS, vol. 6, no. 2, pp. 156–166, 2012.


\bibitem{8}
S. Aviyente, “Compressed sensing framework for EEG compression,” Stat. Signal Process. IEEE, pp. 181–184, 2007.

\bibitem{BioCAS2014}
X. Liu, et al., ``Design of a Net-zero Charge Neural Stimulator with Feedback Control," \emph{BioCAS}, 2014.

\bibitem{9}
D. Gangopadhyay, E. G. Allstot, a M. R. Dixon, K. Natarajan, S. Gupta, and D. J. Allstot, “Compressed Sensing Analog Front-End for Bio-Sensor Applications,” JSSC, vol. 49, no. 2, pp. 426–438, 2014.

\bibitem{12}
X. Liu et al., ``Design of a Closed-Loop, Bidirectional Brain Machine Interface System With Energy Efficient Neural Feature Extraction and PID Control," \emph{TBioCAS}, 2016.

\bibitem{13}
M. Mangia, F. Pareschi, R. Rovatti, and G. Setti, “Adapted Compressed Sensing: A Game Worth Playing,” IEEE Circuits Syst. Mag., vol. 20, no. 1, pp. 40–60, 2020.

\bibitem{15}
Z. Zhang, T. Jung, S. Makeig, and B. Rao, “Compressed sensing of EEG for wireless telemonitoring with low energy consumption and inexpensive hardware,” TBME, pp. 1–4, 2012.

\bibitem{ISCAS2017_SBMI}
X. Liu, et al., ``A Fully Integrated Wireless Sensor-Brain Interface System to Restore Finger Sensation," \emph{ISCAS}, May 2017.

\bibitem{18}
R. Blum, J. Ross, S. Das, E. Brown, and S. Deweerth, “Models of stimulation artifacts applied to integrated circuit design.,” EMBC, vol. 6, pp. 4075–4078, 2004.

\bibitem{ISCAS2017_WM}
X. Liu, et al., ``A Wireless Neuroprosthetic for Augmenting Perception Through Modulated Electrical Stimulation of Somatosensory Cortex," \emph{ISCAS}, May 2017.

\bibitem{Wettels}
N. Wettels, V. J. Santos, R. S. Johansson, and G. E. Loeb, “Biomimetic Tactile Sensor Array,” Adv. Robot., vol. 22, no. 8, pp. 829–849, Jan. 2008.



\end{thebibliography}
\end{document}